\documentstyle[ltwol,epsfig]{article}

\def\be{\begin{equation}}
\def\ee{\end{equation}}
\def\bea{\begin{eqnarray}}
\def\eea{\end{eqnarray}}

\def\ksks{K^0_S K^0_S}
\def\ptsum{\sum{\mathbf{p}_t^*}}
\def\ptss{|\ptsum|^2}
\def\mksks{M_{K_S K_S}}

\begin{document}

\title{
$K^+ K^-$ AND $\ksks$ PRODUCTION IN TWO-PHOTON COLLISIONS AT BELLE}

\author{H. C. HUANG \\ (for the Belle Collaboration)}

\address{
Department of Physics, National Taiwan University, Taipei 106, Taiwan
\\E-mail: hchuang@phys.ntu.edu.tw.}


\twocolumn[\maketitle\abstracts{
The production of a kaon pair ($K^+ K^-$ and $K^0_S K^0_S$) in
two-photon processes has been measured using the Belle detector at
the electron-positron collider KEKB.  We have obtained the
invariant-mass distribution of the processes between 1.3 and 2.3 GeV.
A broad bump structure around 1.7 GeV in the $K^0_S K^0_S$ channel
is confirmed, while, in the $K^+ K^-$ channel, a bump structure is seen
near 1.9 GeV. The angular distribution of the final-state kaons
is analyzed to explore the spin(-helicity) structure in each
invariant-mass region for the two processes.  Contributions
from existing and postulated resonances are discussed.
}]

\section{Introduction}

A high-luminosity electron-positron collider is a good place 
to study 
meson resonances produced by two-photon collisions. 
The well established two-photon resonance in kaon-pair final states 
is the $f'_2(1525)$ meson \cite{oldkk,l3ksks}, 
which is classified as
an almost pure $s\bar{s}$ meson. 
The L3 Experiment at LEP has also reported a resonance-like
peak 
around 1750~MeV in the $\ksks$ final state \cite{l3ksks}.

Meson spectroscopy in the 
1.5 -- 2.0 GeV region is important since 
glueballs are expected to be found in this mass range.
Glueball candidates around 1.7 and 2.2~GeV have been reported in 
$J/\psi \rightarrow \gamma K\bar{K}$ decays by the BES Experiment 
at BEPC \cite{bes}.
Glueball searches are complicated by excitations of
$q\bar{q}$ mesons that also populate the same 
region. Since gluons do not couple to photons, the two-photon partial
decay widths ($\Gamma_{\gamma\gamma}$) of pure glueball 
states are expected to be very small. Two-photon
processes, therefore, play an important role in 
the search for glueballs.
Indeed, null results from the CLEO Experiment at CESR \cite{cleo}
lend some support for the glueball interpretation of the $f_J(2220)$,
while the L3 result suggests the 1700 MeV region should be
studied in further detail.

\section{Experiment}

The analysis is based on data taken by the Belle detector~\cite{belle} 
at the KEKB~\cite{kekb} asymmetric $e^+e^-$ collider in the period
from October, 1999 to December, 2000.  The integrated luminosity of the
data set analyzed for $K^+ K^-$ final startes is 5.86~fb$^{-1}$ and
10.6~fb$^{-1}$ for $\ksks$ final states.  Since the beam-energy
dependence of two-photon processes is very small, we combine 
data taken at the $\Upsilon(4S)$ resonance ($\sqrt{s}= 10.58~{\rm GeV}$)
with data collected at $\sqrt{s}$ values 50 MeV and 60 MeV lower.

The Belle detector~\cite{belle} is a general purpose detector which
includes a 1.5 T 
superconducting solenoid magnet that surrounds the KEKB crossing
point.  Charged tracks are reconstructed using an 50-layer Central
Drift Chamber (CDC) and a 3-layer double-sided Silicon Vertex Detector
(SVD).  Particle identification is accomplished by combining the
responses from a Silica Aerogel \v Cerenkov Counters (ACC) and a Time
of Flight Counter system (TOF) with specific ionization ($dE/dx$)
measurements in the CDC.  A CsI Electromagnetic Calorimeter (ECL)
located inside the solenoid coil is used for photon detection and
electron identification.  Almost all the signal events used in the
analysis were triggered by requiring two or more tracks in the CDC.

\section{$K^+ K^-$ Final States}

\subsection{Event Selection}

Events with only one pair of charged particles 
produced in two-photon processes ($\gamma\gamma \to X^+X^-$) are selected 
with the following criteria:
the scalar sum of track momenta ($\sum |p|$) in an event 
is required to be smaller than 6~GeV/$c$, and the
sum of the calorimeter energies in an event less than 6~GeV.
The event is required to have only one positively charged track and only one 
negatively charged track, where each satisfies the conditions: 
$p_t > 0.4$~GeV/$c$, $|dr| < 1$~cm, $|dz| < 2$~cm, $-0.34 < \cos \theta < 0.82$, 
where $p_t$ is a transverse momentum of a track with respect
to the positron beam axis, and $dr$ and $dz$ are $r$ and $z$ coordinates,
respectively.
All of the above values 
are measured in the laboratory frame. 
%
%
A cut on $p_t$ balance in the $e^+e^-$ 
CM frame,
$|\sum {\bf p}^*_t| < 0.1$~GeV/$c$, is applied to select exclusive-two-track
events from quasi-real two-photon collisions.

We make particle identification cuts for the two tracks 
to select $K^+K^-$ events. 
Since the sample included a large fraction of
$\gamma\gamma \to e^+e^-$ events, 
we reject them by
requiring $E/p < 0.8$ for the two tracks,
where $E/p$ is the ratio of the energy deposit on ECL to the momentum.  
Charged kaons are selected using
TOF and ACC information
with the criteria that the likelihood ratios 
for $K/\pi$ and $K/p$ separation are larger than 0.8. 
In addition, $dE/dx$ information from CDC is also used
for identifying kaons in the lowest invariant mass region.

\subsection{Invariant Mass Distribution}

\begin{figure}
\center
\epsfig{file=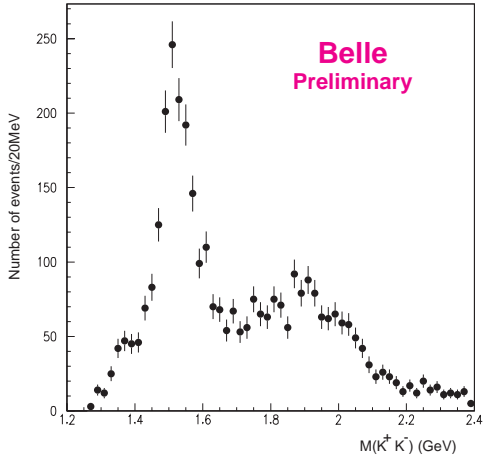,width=2.5in}
\caption{The invariant mass distribution for $\gamma \gamma \to K^+
K^-$ events.}
\label{fg:kkm1}
\end{figure}

After the application of all the selection criteria,
3674 events remain.  Figure~\ref{fg:kkm1} shows the $K^+ K^-$
invariant mass distribution for the selected samples.  The peak at
$1.48 - 1.56$ GeV is from 
$f'_2(1525) \to K^+K^-$.  In addition, there is a broad bump in the
$1.7 - 2.1$ GeV region. 
No significant enhancement around 2.23 GeV is observed. 

\begin{figure}
\center
\epsfig{file=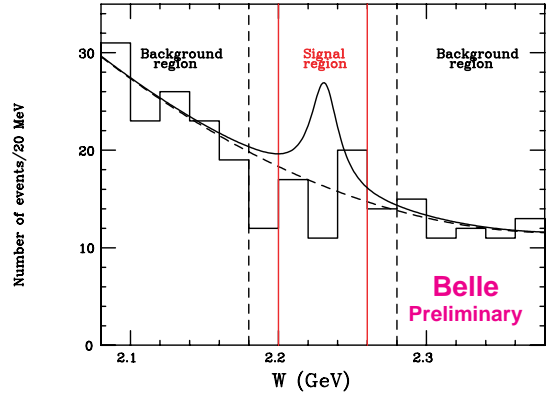,width=2.8in}
\caption{$K^+ K^-$ mass distribution (GeV) observed in data near the
$f_J(2220)$ mass.  The background shape (dashed line) is assumed to be
a second-order polynomial.  The solid line is the sum of a fit to the
background and the signal line shape for central values of the
resonance parameters, $M_{f_J} = 2.231$ GeV and $\Gamma_{f_J} = 23$
MeV, corresponding to the observed $95\%$ C.L. upper limit of 14.7
excess events.}
\label{fg:kkfj}
\end{figure}

The mass region near the $f_J(2220)$ is shown in detail in 
Figure~\ref{fg:kkfj}.
The signal region is chosen to be within 2220 MeV and 2260 MeV.  To
obtain the background shape, we fit the $K^+ K^-$ mass distribution as a
second-order polynomial from the sideband regions.  
We observed 48 events in the signal region and the expected background
is 49.2.  Using a Poisson distribution with background,
we obtain a $95\%$ C.L. upper limit 
\[
	\Gamma_{\gamma\gamma}(f_J(2220)) \times
	{\mathcal{B}}(f_J(2220) \to K^+ K^-) < 
	1.5 \mathrm{eV}
\]
under the assumption of $(J, \lambda) = (2, 2)$.

\subsection{Analysis of the Angular Distribution}
\label{sc:pwa}

We made a model-independent partial-wave analysis for
the angular distribution of the $K^+K^-$ events.  
For the total angular momentum ($J$) of the $\gamma\gamma$ system, we
assumed that partial waves with $J \ge 4$ were negligible. 
The odd-$J$ waves are forbidden by the symmetric nature 
of the initial state and the parity conservation.
In the $\gamma\gamma$ system, the differential cross 
section can be written as

\begin{eqnarray}
\frac{d \sigma}{d \Omega} = |{\cal H}(J=0, \lambda=0) +
{\cal H}(J=2, \lambda=0)|^2 \nonumber \\
+ |{\cal H}(J=2, \lambda=2)|^2 ~~~~~~~~~~~~~~~~~
\label{eq1}
\end{eqnarray}

\noindent
where the ${\cal H}$'s are the partial-wave amplitudes for the $J$ 
and  $\lambda$ (helicity) states.

Equation~(\ref{eq1}) includes four real-number parameters, the sizes of the
three partial wave amplitudes and one relative phase between
the two $\lambda=0$ waves.  Expressing Eq.~(\ref{eq1}) with
$d$-functions, the angular distribution can be reduced to

\begin{eqnarray}
\frac{d \sigma}{d|\cos \theta^*|} = {\cal X} 
+ {\cal Y}|d^2_{00}(\cos \theta^*)|^2 
+ {\cal Z}|d^2_{20}(\cos \theta^*)|^2 
\label{eq2}
\end{eqnarray}

\noindent
Here the three parameters ${\cal X}$, ${\cal Y}$ and
${\cal Z}$ can be written in terms of the three partial-wave
amplitudes and one relative phase, as explicitly presented in 
the Reference~\cite{yabuki}.
However, in Eq.~(\ref{eq2}), 
we still cannot determine ${\cal Y}$  and ${\cal Z}$ independently, 
since the two $d$-functions have a
similar shape for $|\cos \theta^*| <0.5$.  Therefore, we introduced
two new parameters, ${\cal Z}'={\cal Y}+{\cal Z}$ and $f={\cal Y}/{\cal Z}$,
and assumed $f=0.$5.  The fitting results for ${\cal X}$ and 
${\cal Z}'$ are not sensitive to
the choice of the $f$ value.  
The parameter ${\cal X}$ contains the $J=0$ cross
section dominantly, whereas the parameter ${\cal Z}'$ is
dominated by the $J=2$ contributions, if the contribution
of the interference effect between the two $\lambda=0$ states 
are sufficiently small.
 
\begin{figure}
\center
\epsfig{file=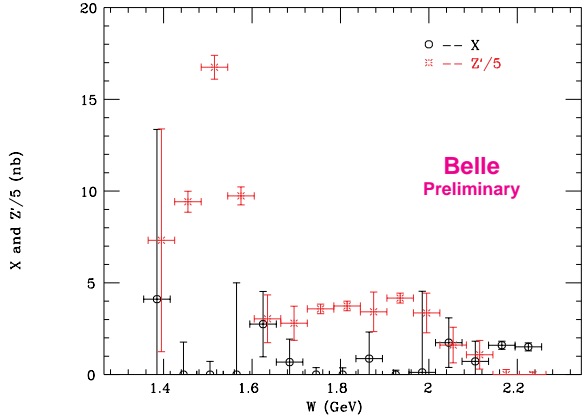,width=3in}
\caption{The contribution to the total cross section
from the components ${\cal X}$ (dominated by $J=0$, 
open circles) and ${\cal Z}'/5$
(dominated by $J=2$, asterisk marks). }
\label{fg:kkang}
\end{figure}

We have fit the ${\cal X}$ and ${\cal Z}'$ values at
different $W$ points independently. Figure~\ref{fg:kkang} shows the
$W$ dependences of the parameters (we give the results in
${\cal Z}'/5$ to normalize them to the contribution
in the total cross section).  It is clear that the two peak structures from
$f_2'(1525)$ and of 1.9~GeV region are dominated by $J=2$ components.
The $J=0$ component is consistent with zero except in the highest 
$W$ region.

\section{$\ksks$ Final States}

\subsection{Event Selection}

Only events where both $K^0_S$'s decay to $\pi^+\pi^-$ are considered.
To select events in which the $K^0_S$'s were produced in two-photon
collisions, we require that: the scalar sum of track momenta ($\sum
{ |p| }$) in an event is smaller than 6 GeV; 
the sum of calorimeter energies in an event 
is smaller than 6 GeV; 
there are exactly four reconstructed charged tracks with a net
charge of zero; 
and the total momentum imbalance in the transverse plane  satisfies 
$P_t^2 \equiv \ptss < 0.1 \mathrm{GeV}^2$, to select events with small
incident photon virtuality.

The charged tracks are assumed to be pions.  A pair of
oppositely
charged tracks are combined to form a $K^0_S$ candidate if their
reconstructed vertex is displaced from the primary interaction vertex.
The tracks are then refit with the constraint that they
come from the reconstructed vertex.  They are kept if the resulting
invariant mass for the pair is within 10 MeV of $M_{K^0}$ \cite{pdg}
and the combined momentum vector points back to the interaction
vertex.  After these selection cuts, the mass resolution is obtained
to be $2.6 \ \mathrm{ MeV}$.  The pairs are again fit, constraining
the mass of each $K^0_S$ candidate to 
the nominal PDG value \cite{pdg}.  

\begin{figure}
\begin{center}
\epsfig{file=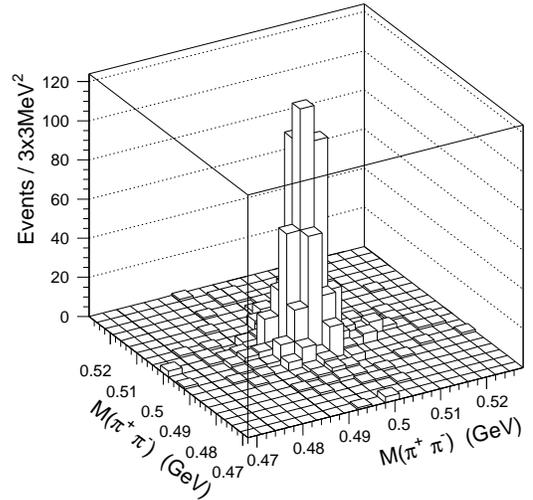, height=3.1in}
\caption{$m_{\pi^+\pi^-}$ of one $K_S^0$ candidate versus
$m_{\pi^+\pi^-}$ of the other.  There is a strong enhancement near
the $(m_{K_S^0}, m_{K_S^0})$ point over a very small background.
}
\label{fg:ksks}
\end{center}
\end{figure}

Since the two $K^0_S$'s are
produced back-to-back 
in the transverse plane, the angle between the flight directions of
the two $K^0_S$ candidates in this plane is required to be larger than
$160^\circ$.  A legoplot of the two unconstrained pion-pair masses is shown in
Figure~\ref{fg:ksks}.  There is a strong enhancement at 
$(m_{K_S^0}, m_{K_S^0})$ and the background fraction is very low.
After applying all the selection cuts, 1511 events are found in the data 
sample.

\subsection{Analysis}

\begin{figure}[ht]
\begin{center}
\epsfig{file=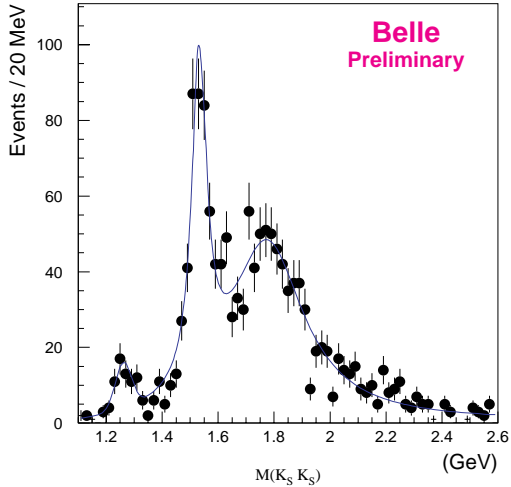, height=3in}
\caption{The $K^0_S K^0_S$ invariant mass spectrum for the
$\gamma\gamma \to \ksks$ process.  The solid line is a fit with two
Breit-Wigner functions for the $f'_2(1525)$ and 1750 MeV region, and a
Gaussian for the $f_2(1270)-a^0_2(1320)$ region 
plus a constant background.  See the text for details.}
\label{fg:massksks}
\end{center}
\end{figure}

%
%
The resulting $\ksks$ invariant mass spectrum is shown in
Figure~\ref{fg:massksks}.  The spectrum obtained is similar to the one 
reported by 
the L3 Experiment \cite{l3ksks}.  The spectrum is dominated by the 
$f'_2(1525)$ resonance and a clear enhancement is visible in the 1750
MeV region.  The $f_2(1270)-a^0_2(1320)$ region shows the
destructive  interference expected in the $\ksks$ final
state \cite{lipkin}.  The event rate in this region is suppressed by
the detection efficiency.  

The $\ksks$ invariant mass spectrum is fitted by minimizing a $\chi^2$
function for the expected value in the $i$th bin.
\begin{equation}
E_i = \xi_i \cdot ( G_i + BW^{(1)}_i + BW^{(2)}_i + B_i) \, .
\end{equation}
where $\xi_i$ is the detection efficiency parameterized from the
Monte Carlo.  The small bump in the $f_2(1270)-a^0_2(1320)$ region is
described by a Gaussian function $G_i$ with its mean and sigma as free
parameters.  The peaks at $f'_2(1525)$ and 1750 MeV regions are
described by Breit-Wigner functions $BW_i$ with their masses and widths
taken as free parameters. 
The background
is assumed to be a constant ($B_i$) in the fit.
The result of the fit is
illustrated by the curve in Figure~\ref{fg:massksks}.  The parameters
obtained are summarized in Table~\ref{tb:massfit}.

\begin{table}
\caption{Results of the fit to the $\ksks$ mass spectrum.} 
\label{tb:massfit}
\begin{center}
\begin{tabular}{|l|c|c|c|}
\hline
		& $f'_2(1525)$	& 1750 MeV	\\
\hline
Mass (MeV)	& $1532\pm4$	& $1768\pm9.6$	\\
\hline
Width (MeV)	& $64\pm6.8$	& $323\pm29$	\\
\hline
No. of Events	& $414\pm36$	& $967\pm72$	\\
\hline
\end{tabular}
\end{center}
\end{table}

\subsection{The $f_J(2220)$ Mass Region}


No enhancement at the $f_J(2220)$ mass is observed.  
We assume a mass of 2230 MeV and a total width of 20 MeV.  The signal
region is chosen to be $\pm 40 \mathrm{~MeV}$ around 2230 MeV.  To
obtain a background shape, we fit the $\mksks$ distribution with a
linear function from 2.11 to 2.35 GeV, excluding the signal region.
We observed 36 events in the signal region and the expected background
is 27.0.  Using a Poisson distribution with background,
we obtain an upper limit of 20.7 signal events at the $95\%$ C.L.
Assuming $(J,\lambda) = (2,2)$ for $f_J(2220)$, this corresponds to 
\[
	\Gamma_{\gamma\gamma}(f_J(2220)) \times
	{\mathcal{B}}(f_J(2220) \to \ksks) < 
	1.17 \mathrm{eV}
\]
at $95\%$ C.L. without considering the systematic errors.

\subsection{Angular Analysis}

\begin{figure}
\center
\epsfig{file=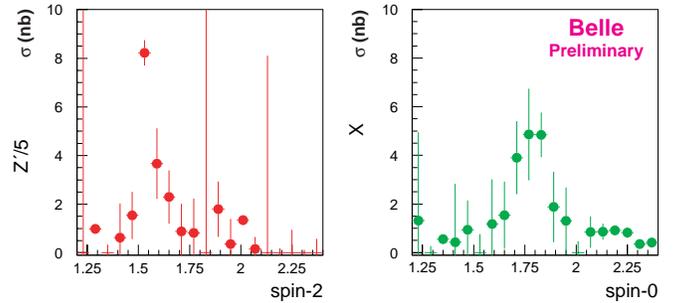,width=3.4in}
\caption{The contribution to the total cross section from the
components (a) ${\mathcal{Z}}' / 5$, dominated by spin-2 component, and (b)
$\mathcal{X}$, dominated by spin-0 component.}
\label{fg:ksksang}
\end{figure}

We have applied the same method of partial wave analysis, described in 
Sec.~\ref{sc:pwa}, to the $\ksks$ data sample.  The angular
distribution of each $W$ bin is fit by Eq.~\ref{eq2}.
Figure~\ref{fg:ksksang} shows the $W$ dependences of the fitted
results, ${\cal Z}' / 5$ and $\mathcal{X}$, which correspond to the
contribution of $J = 2$ and $J = 0$ components in the total cross
section, respectively.  It is clear that the peak of $f'_2(1525)$ is
dominated by spin-2 component as expected.  However, there is a large
spin-0 component in the 1750 MeV region while some spin-2 structure
appears around $1.8 - 2.0$ GeV.

\section{Summary}

We have studied the reactions $\gamma\gamma\rightarrow K^+ K^-$ and $\gamma\gamma\rightarrow \ksks$ 
using the large data samples collected by the Belle experiment
at KEKB.  A prominent $f'_2(1525)$ resonance is observed in both channels.
A broad structure in the $1.7 - 2.1$ GeV region is found in $K^+K^-$
and an enhancement around 1750 MeV is observed in $\ksks$.  A partial
wave analysis has been performed to explore these resonance
structures.  Upper limits for $f_J(2220)$ are also obtained in both
channels, respectively.

\end{document}